\documentclass[a4paper]{jpconf}

\usepackage[utf8]{inputenc}
\usepackage[T1]{fontenc}
\usepackage{lmodern}
\usepackage[english]{babel}
\usepackage{textcomp}

\usepackage{hyperref}

\usepackage{graphicx}

\usepackage{amsmath,amsfonts,amssymb}
\usepackage{slashed,stackrel}

\usepackage{feynmp}
\def\fmfstyle{
  \fmfset{thin}{0.3pt}         
  \fmfset{thick}{1.5thin}      
  \fmfset{arrow_ang}{15}       
  \fmfset{arrow_len}{2.0mm}    
  \fmfset{curly_len}{1.5mm}    
  \fmfset{dash_len}{2.0mm}     
  \fmfset{dot_len}{1.0mm}      
  \fmfset{dot_size}{1.0mm}     
  \fmfset{zigzag_len}{2.0mm}   
  \fmfset{zigzag_width}{0.2mm} 
}

\newcommand{\MSbar}{\ensuremath{{\overline{\text{MS}}}}}

\renewcommand{\d}[2][D]{\ensuremath{\operatorname{d}^{#1}\!{#2}}}
\newcommand{\p}[1][]{\ensuremath{\partial{#1}}}

\newcommand{\myset}[1]{\ensuremath{\left\{ #1 \right\}}}
\newcommand{\myvec}[1]{\ensuremath{\left( #1 \right)}}

\newcommand{\trans}[1]{#1^{\operatorname{T}}}

\makeatletter
\renewcommand*\env@matrix[1][*\c@MaxMatrixCols c]{%
  \hskip -\arraycolsep
  \let\@ifnextchar\new@ifnextchar
  \array{#1}}
\makeatother

\begin{document}

\begin{flushright}
  DESY 16-036
\end{flushright}

\title{The {\tt Mathematica} package {\tt TopoID} and its application to
  the Higgs boson production cross section}

\author{Jens Hoff}

\address{Deutsches Elektronen-Synchrotron~(DESY), Platanenallee 6,
  D-15738 Zeuthen, DE}

\ead{jens.hoff@desy.de}

\begin{abstract}
  We present the {\tt Mathematica} package {\tt TopoID} which aims at
  the automation of several steps in multiloop calculations.  The
  algorithm which lies at the very core of the package is described and
  illustrated with an example.  The main features of {\tt TopoID} are
  stated and some of them are briefly demonstrated for NLO or NNLO Higgs
  boson production.
\end{abstract}

\section{Introduction}

With the LHC running in the coming years at $14\,\text{TeV}$, the
experimental precision for observables will further improve. Likewise,
the theoretical uncertainty needs to be decreased which means higher
orders in the perturbation series need to be computed.  Some quantities
under consideration presently are for example the Higgs boson production
cross section at N$^3$LO~\cite{Anastasiou:2014lda,Anzai:2015wma}, the
four-loop \MSbar-OS relation~\cite{Marquard:2015qpa,Marquard:2016dcn},
the four-loop cusp anomalous dimension~\cite{Henn:2016men} or the
five-loop $\beta$-function~\cite{Baikov:2016tgj}.  All these
calculations are based on the evaluation of Feynman diagrams which
usually suffers from a factorial growth of complexity with higher
orders.

For most of the necessary steps in such a calculation various program
packages are already available publically.  Color, Lorentz and Dirac
algebra (tensorial structures) can be handled, e.g., with {\tt
  FORM}~\cite{Kuipers:2012rf}, {\tt FeynCalc}~\cite{Shtabovenko:2016sxi}
or {\tt HEPMath}~\cite{Wiebusch:2014qba}.  The resulting scalar
integrals can then be classified in so-called Feynman integral families
or ``topologies''.  {\tt TopoID}~\cite{Grigo:2014oqa,diss_Hoff} is a
package dedicated especially to this task (and others).  The integrals
within each topology are subject to integration-by-parts
identities~(IBPs)~\cite{Chetyrkin:1981qh} which allow for a systematic
reduction to ``master integrals''.  That is, a large set of integrals
can be expressed as linear combinations of only relatively few basis
integrals.  Public reduction programs include {\tt
  MINCER}~\cite{Gorishnii:1989gt} for three-loop massless propagtors,
{\tt MATAD}~\cite{Steinhauser:2000ry} for three-loop massive tadpoles or
{\tt AIR}~\cite{Anastasiou:2004vj}, {\tt FIRE}~\cite{Smirnov:2014hma}
and {\tt Reduze}~\cite{vonManteuffel:2012np} implementing Laporta's
approach~\cite{Laporta:2001dd} for generic topologies.

{\tt TopoID} aims at taming the diagrammatic or ``topologic'' complexity
that arises in a calculation.  The complexity involved in algebraic
operations on the coefficients of integrals in reduction relations or in
the function classes for master integrals, however, is a different and
still open subject.

\subsection{Example: Higgs boson production at NLO}
\label{subs:NLO}

Let us illustrate the concepts of topology and integral reduction by the
single topology needed for the NLO corrections to Higgs boson production
due to real radiation:
{\small
  \begin{align}
    & \begin{fmffile}{dias/NLO_top}
      \parbox{100pt}{
        \fmfframe(0,5)(0,5){
          \begin{fmfgraph*}(100,50)
            \fmfstyle
            \fmfleft{i2,i1}
            \fmfright{o4,o3}
            \fmf{plain_arrow,label=$p_1$,l.d=4.0}{i1,v1}
            \fmf{plain_arrow,label=$p_2$,l.d=4.0,l.s=left}{i2,v2}
            \fmf{plain_arrow,label=$p_1$,l.d=4.0}{v3,o3}
            \fmf{plain_arrow,label=$p_2$,l.d=4.0,l.s=left}{v4,o4}
            \fmf{phantom,t=0.3}{v1,v3}
            \fmf{phantom,t=0.3}{v2,v4}
            \fmffreeze
            \fmf{dbl_plain_arrow,rubout,t=0.3,label=$d_1$,l.d=4.0,l.s=right}{v3,v5}
            \fmf{dbl_plain,rubout,t=0.3}{v5,v2}
            \fmf{plain_arrow,t=0.3,label=$d_2$,l.d=4.0,l.s=left}{v1,v6}
            \fmf{plain,t=0.3}{v6,v4}
            \fmf{plain_arrow,t=0.0,label=$d_3$,l.d=4.0,l.s=left}{v2,v1}
            \fmf{plain_arrow,t=0.0,label=$d_4$,l.d=4.0}{v4,v3}
          \end{fmfgraph*}}}
    \end{fmffile}
    = \int \d[D]{k_1} \:
    \frac{1}{d_1^{a_1} d_2^{a_2} d_3^{a_3}d_4^{a_4}}
    = T\!\left( a_1, a_2, a_3, a_4 \right) \quad \text{with} \quad
    \begin{aligned}
      d_1 &= m_H^2 + k_1^2,\\
      d_2 &= (p_1 + p_2 + k_1)^2,\\
      d_3 &= (p_2 + k_1)^2,\\
      d_4 &= (p_1 + k_1)^2.
    \end{aligned}
    \label{eq:top}
  \end{align}}
The non-planar box topology $T$ in Eq.~\eqref{eq:top} consists of four
scalar propagators $1/d_1, \ldots, 1/d_4$ raised to generic integer
powers $a_1, \ldots, a_4$ ($a_i = 0$ means contraction of line $i$).
Note, this topology is defined in forward scattering kinematics to
faciliate the use of the optical theorem (for a detailed discussion see
Ref.~\cite{diss_Hoff}).  Forward scattering means outgoing momenta are
fixed by incoming ones ($p_3 = p_1$ and $p_4 = p_2$) and in this sense
above topology is non-planar.  The double line indicates the Higgs
boson, arrows show the momentum flow.  Due to the linear dependence of
$d_1, \ldots, d_4$ (with respect to the scalar products $p_1 \cdot k_1$,
$p_2 \cdot k_1$ and $k_1^2$) and since integrals with $d_1$ or $d_2$
eliminated do not contribute via the optical theorem, partial
fractioning allows to eliminate either $d_3$ or $d_4$ in all appearing
expressions.  This leaves us with two symmetric triangle topologies
$T\!\left( a_1, a_2, a_3, 0 \right) = T\!\left( a_1, a_2, 0, a_3
\right)$.  Gauss' theorem in $D$ dimensions gives in this case rise to
three IBPs, one for the contraction with either the loop momentum $k_1$
or one of the external momenta $p_1$ and $p_2$:
\begin{align}
  & 0 = \int \d[D]{k_1} \: \frac{\p{}}{\p{k_1^\mu}}
  \left\{ k_1^\mu, p_1^\mu, p_2^\mu \right\}
  \frac{1}{d_1^{a_1} d_2^{a_2} d_3^{a_3}}.
  \label{eq:ibp}
\end{align}
The relations obtained from Eq.~\eqref{eq:ibp} by fixing $a_1, \ldots,
a_3$ to integer values then involve integrals with differing denominator
powers.  Generating a set of such relations for all the needed integrals
of a family gives a linear system which can be solved in terms of master
integrals.  In this case only a single master integral (a bubble with
$a_1 = a_2 = 1$, $a_3 = a_4 = 0$) emerges in the end.

\section{Canonical ordering}

The most important capability of {\tt TopoID} is the fast and efficient
{\bf ID}entification of isomorphic {\bf Topo}logies.  This is essential
since the routing of loop momenta and the order of propagators in the
definition of a topology are ambiguous.  The basic idea is to use the
sum $\mathcal{U} + \mathcal{F}$ of the Symanzik polynomials appearing in
the Feynman representation, which is independent of loop momenta, as an
identifier.  Therefore, the Feynman parameters $\myset{\alpha_j}$
(corresponding to scalar propagators) are permuted in a unique
algorithmic way. See also Ref.~\cite{Pak:2011xt}.

The task of the algorithm can be stated without reference to Feynman
integrals: bring the polynomial $P$ with $m$ terms into a unique form
$\hat{P}$ by renaming the $n$ variables $\myset{x_j}$.  This can be
achieved by the following steps:
\begin{enumerate}
\item \label{enum:1} Convert $P$ into a $m \times (n + 1)$ matrix
  $M^{(0)}$.  Each row corresponds to a term, the first column to
  constant coefficients and the remaining columns to powers of the
  $\myset{x_j}$ in the monomial.
\item \label{enum:2} Start with the above $M^{(0)}$ and in the second
  column ($k = 1$).
\item \label{enum:3} Compute for all considered matrices $M^{(k),
    \sigma}$ all transpositions of columns $k$ and $k + 1, \ldots, n$
  where the index $\sigma$ collects all applied permutations.
\item \label{enum:4} Sort rows in each matrix lexicographically by the
  first $k$ columns.
\item \label{enum:5} Extract the lexicographically largest vector from
  columns $k$ of all matrices.
\item \label{enum:6} Keep only matrices with this maximal vector.  If $k
  < n - 1$, increment $k$ and go to step (\ref{enum:3}).
\item \label{enum:7} Each remaining matrix encodes the same unique
  $\hat{P}_{\sigma}$ and a corresponding permutation of variables $\sigma$.
\end{enumerate}

\subsection{Example: a simple polynomial in two variables}

We demonstrate the above algorithm in a simple example where equation
labels indicate the respective step and iteration:
{\small
  \begin{align}
    & P = x_1^2 + 2 x_1 x_2 + x_2^2 + x_3^2 \; \to \; M^{(0)} =
    \begin{pmatrix}
      1 & 2 & 0 & 0\\
      2 & 1 & 1 & 0\\
      1 & 0 & 2 & 0\\
      1 & 0 & 0 & 2
    \end{pmatrix}.
    \tag{\ref{enum:1}}\\
    & S^{(1)} = \myset{M^{(0) (123)} = M^{(0)}}, \; k = 1.
    \tag{\ref{enum:2}}\\
    & \begin{aligned}
      S^{\prime (1)}: \; &
      M^{\prime (1) (123)} =
      \begin{pmatrix}[cc|cc]
        1 & 2 & 0 & 0\\
        2 & 1 & 1 & 0\\
        1 & 0 & 2 & 0\\
        1 & 0 & 0 & 2
      \end{pmatrix}, \;
      M^{\prime (1) (213)} =
      \begin{pmatrix}[cc|cc]
        1 & 0 & 2 & 0\\
        2 & 1 & 1 & 0\\
        1 & 2 & 0 & 0\\
        1 & 0 & 0 & 2
      \end{pmatrix}, \;
      M^{\prime (1) (321)} =
      \begin{pmatrix}[cc|cc]
        1 & 0 & 0 & 2\\
        2 & 0 & 1 & 1\\
        1 & 0 & 2 & 0\\
        1 & 2 & 0 & 0
      \end{pmatrix}.
    \end{aligned}
    \tag{\ref{enum:3}-1}\\
    & \begin{aligned}
      S^{\prime \prime (1)}: \; &
      M^{\prime \prime (1) (123)} =
      \begin{pmatrix}
        1 & \bf{0} & 0 & 2\\
        1 & \bf{0} & 2 & 0\\
        1 & \bf{2} & 0 & 0\\
        2 & \bf{1} & 1 & 0
      \end{pmatrix}, \;
      M^{\prime \prime (1) (213)} =
      \begin{pmatrix}
        1 & \bf{0} & 2 & 0\\
        1 & \bf{0} & 0 & 2\\
        1 & \bf{2} & 0 & 0\\
        2 & \bf{1} & 1 & 0
      \end{pmatrix}, \;
      M^{\prime \prime (1) (321)} =
      \begin{pmatrix}
        1 & 0 & 0 & 2\\
        1 & 0 & 2 & 0\\
        1 & 2 & 0 & 0\\
        2 & 0 & 1 & 1
      \end{pmatrix}.
    \end{aligned}
    \tag{\ref{enum:4}-1}\\
    & \hat{M}^{\prime \prime (1)} = \trans{\myvec{0, 0, 2, 1}}.
    \tag{\ref{enum:5}-1}\\
    & S^{(2)} = \myset{
      M^{\prime \prime (1) (123)}, M^{\prime \prime (1) (213)}
    }, \; k = 2.
    \tag{\ref{enum:6}-1}\\
    & \begin{aligned}
      S^{\prime (2)}: \; &
      M^{\prime (2) (123)} =
      \begin{pmatrix}[ccc|c]
        1 & 0 & 0 & 2\\
        1 & 0 & 2 & 0\\
        1 & 2 & 0 & 0\\
        2 & 1 & 1 & 0
      \end{pmatrix}, \;
      M^{\prime (2) (132)} =
      \begin{pmatrix}[ccc|c]
        1 & 0 & 2 & 0\\
        1 & 0 & 0 & 2\\
        1 & 2 & 0 & 0\\
        2 & 1 & 0 & 1
      \end{pmatrix},\\
      & M^{\prime (2) (213)} =
      \begin{pmatrix}[ccc|c]
        1 & 0 & 2 & 0\\
        1 & 0 & 0 & 2\\
        1 & 2 & 0 & 0\\
        2 & 1 & 1 & 0
      \end{pmatrix}, \;
      M^{\prime (2) (231)} =
      \begin{pmatrix}[ccc|c]
        1 & 0 & 0 & 2\\
        1 & 0 & 2 & 0\\
        1 & 2 & 0 & 0\\
        2 & 1 & 0 & 1
      \end{pmatrix}.
    \end{aligned}
    \tag{\ref{enum:3}-2}\\
    & \begin{aligned}
      S^{\prime \prime (2)}: \; &
      M^{\prime \prime (2) (123)} =
      \begin{pmatrix}
        1 & 0 & \bf{0} & 2\\
        1 & 0 & \bf{2} & 0\\
        1 & 2 & \bf{0} & 0\\
        2 & 1 & \bf{1} & 0
      \end{pmatrix}, \;
      M^{\prime \prime (2) (132)} =
      \begin{pmatrix}
        1 & 0 & 0 & 2\\
        1 & 0 & 2 & 0\\
        1 & 2 & 0 & 0\\
        2 & 1 & 0 & 1
      \end{pmatrix},\\
      & M^{\prime \prime (2) (213)} =
      \begin{pmatrix}
        1 & 0 & \bf{0} & 2\\
        1 & 0 & \bf{2} & 0\\
        1 & 2 & \bf{0} & 0\\
        2 & 1 & \bf{1} & 0
      \end{pmatrix}, \;
      M^{\prime \prime (2) (231)} =
      \begin{pmatrix}
        1 & 0 & 0 & 2\\
        1 & 0 & 2 & 0\\
        1 & 2 & 0 & 0\\
        2 & 1 & 0 & 1
      \end{pmatrix}.
    \end{aligned}
    \tag{\ref{enum:4}-2}\\
    & \hat{M}^{\prime \prime (2)} = \trans{\myvec{0, 2, 0, 1}}.
    \tag{\ref{enum:5}-2}\\
    & S^{(2)} = \myset{
      M^{\prime \prime (2) (123)}, M^{\prime \prime (2) (213)}
    }.
    \tag{\ref{enum:6}-2}\\
    & \hat{P} = P = x_3^2 + x_2^2 + x_1^2 + 2 x_1 x_2 \, , \;
    \hat{\sigma} = \myset{(123), (213)}.
    \tag{\ref{enum:7}}
  \end{align}}
We observe, $P$ was already in its unique form $\hat{P}$ (apart from the
order of terms) and the algorithm resulted in two permutations
$\myvec{123}$ and $\myvec{213}$ which correspond to the apparent
symmetry in exchanging $x_1$ and $x_2$.  In the context of Feynman
integrals this means the algorithm gives not just a unique identifier
$\hat{\mathcal{U}} + \hat{\mathcal{F}}$ but also its symmetries.

\section{Features of {\tt TopoID}}

\begin{figure}
  \begin{center}
    \parbox{65pt}{
      \centering
      \includegraphics[scale=0.65,angle=180]{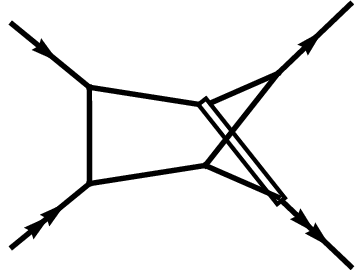}}
    \hfill
    \parbox{65pt}{
      \centering
      \includegraphics[scale=0.65,angle=180]{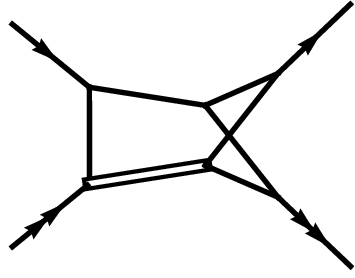}}
    \hfill
    \parbox{65pt}{
      \centering
      \includegraphics[scale=0.65,angle=180]{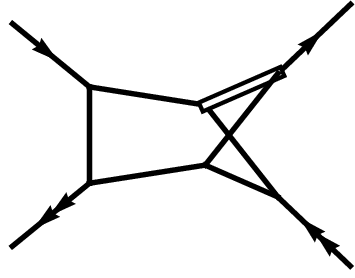}}
    \hfill
    \parbox{65pt}{
      \centering
      \includegraphics[scale=0.65,angle=180]{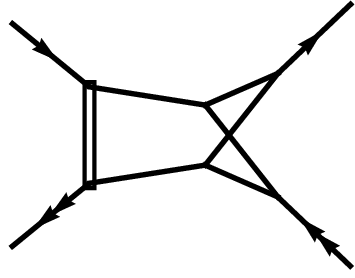}}
    \hfill
    \parbox{65pt}{
      \centering
      \includegraphics[scale=0.65,angle=180]{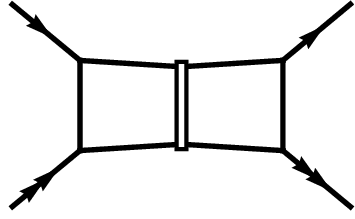}}\\[1.0pc]
    \parbox{65pt}{
      \centering
      \includegraphics[scale=0.65,angle=180]{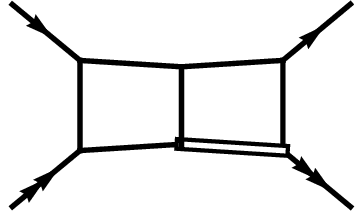}}
    \hfill
    \parbox{65pt}{
      \centering
      \includegraphics[scale=0.65,angle=180]{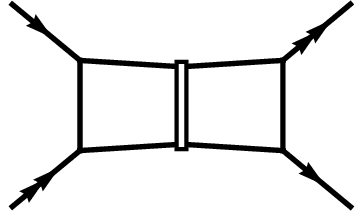}}
    \hfill
    \parbox{65pt}{
      \centering
      \includegraphics[scale=0.65,angle=180]{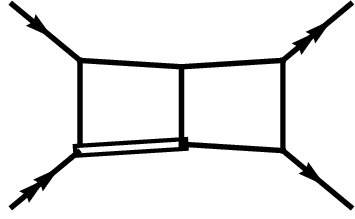}}
    \hfill
    \reflectbox{
      \parbox{40pt}{
        \centering
        \includegraphics[scale=0.7,angle=90]{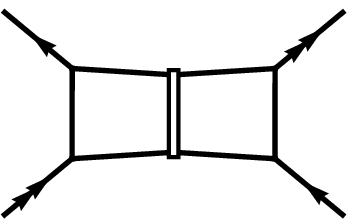}}}
    \hfill
    \reflectbox{
      \parbox{40pt}{
        \centering
        \includegraphics[scale=0.7,angle=90]{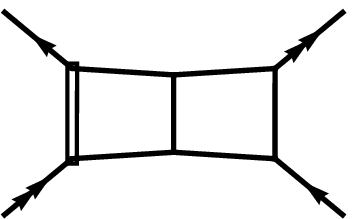}}}
    \hspace*{0.5pc}
    \parbox{65pt}{
      \centering
      \includegraphics[scale=0.7,angle=180]{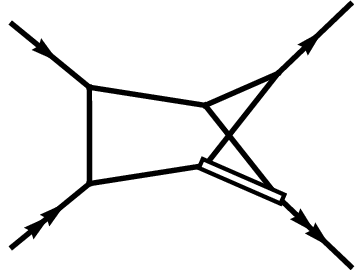}}
  \end{center}
  \caption{Minimal set of topologies needed for the real corrections to
    the Higgs boson production cross section at NNLO.  This set is
    sufficient for the calculation of all $2\,946$ appearing diagrams.}
  \label{fig:tops}
\end{figure}

{\tt TopoID} is written in {\tt Mathematica} and assists in obtaining
results for the amplitude of a given process in terms of a minimal set
of master integrals.  However, for the computationally expensive parts
of a calculation, much faster {\tt FORM} code is generated that does not
demand for a {\tt Mathematica} license.  The main input for this task
are the Feynman diagrams which can be generated, e.g., by {\tt
  QGRAF}~\cite{Nogueira:1991ex}.

Starting from the diagrams, {\tt TopoID} identifies a minimal set of
topologies.  That is, by contraction and permutation of propagators all
scalar integrals arising from the diagrams can be mapped to at least one
of the topologies in the minimal set.  See Fig.~\ref{fig:tops} for a
minimal set sufficient for the real contributions to NNLO Higgs boson
production.

In case the topologies identified in a first step expose linearly
dependent propagators, see Sec.~\ref{subs:NLO}, linear independent
subtopologies are detected and partial fractioning relations are
generated automatically.  The algorithm described in
Ref.~\cite{Pak:2011xt} which employs Gr\"obner bases is used.  For
Eq.~\ref{eq:top} the following set of rules is generated whose repeated
application is guaranteed to terminate:
{\small
  \begin{align}
    & \begin{aligned}
      d_4 & \to -m_H^2 + s + d_1 + d_2 - d_3,\\
      d_3 / d_4 & \to
      \left( -m_H^2 + s + d_1 + d_2 - d_4 \right) /
      d_4,\\
      d_2 / \left( d_3 d_4 \right) & \to
      \left( m_H^2 - s - d_1 + d_3 + d_4 \right) /
      \left( d_3 d_4 \right),\\
      d_1 / \left( d_2 d_3 d_4 \right) & \to
      \left( m_H^2 - s - d_2 + d_3 + d_4 \right) /
      \left( d_2 d_3 d_4 \right),\\
      1 / \left( d_1 d_2 d_3 d_4 \right) & \to
      \left( d_1 + d_2 - d_3 - d_4 \right) /
      \left( \left( m_H^2 - s \right) d_1 d_2 d_3 d_4 \right).
    \end{aligned}
  \end{align}}

It is of course understood that {\tt TopoID} is capable of dealing with
various properties of topolgies, such as completeness with respect to
scalar products involving loop momenta, distinct and scaleless
subtopologies and symmetries.  Also graphs corresponding to a set of
propagators can be reconstructed and unitarity cuts (used extensively in
connection with the optical theorem and forward scattering) can be
revealed in a very efficient way.

Finally, let us mention that also relations among master integrals
(emerging from different topologies) can be found using {\tt TopoID}
(usually in conjunction with a Laporta algorithm).  These relations
emerge by identifying equivalent integrand denominators and transforming
possible numerators accordingly.  Such relations can simplify a
calculation tremendously or are very useful cross-checks.  An example
for such a relation from NNLO Higgs boson production can be sketched as
follows:
\begin{align}
  \begin{fmffile}{dias/NNLO_rel}
    \parbox{40pt}{
      \begin{fmfgraph*}(40,30)
        \fmfstyle
        \fmfcurved
        \fmfforce{(0.7w,-0.05h)}{c1}
        \fmfforce{(0.5w,1.05h)}{c2}
        \fmf{zigzag,w=2.0,f=(0.7,,0.7,,0.7)}{c1,c2}
        \fmfleft{i2,i1}
        \fmfright{o2,o1}
        \fmf{plain_arrow,t=4.0}{i1,v1}
        \fmf{plain,t=4.0}{i2,v1}
        \fmf{plain_arrow,t=4.0}{v2,o1}
        \fmf{plain,t=4.0}{v3,o2}
        \fmf{dbl_plain,t=0.0,rubout}{v2,m1}
        \fmf{plain,rubout}{v1,v2}
        \fmf{plain,t=2.0,rubout}{v1,m1}
        \fmf{plain,rubout,t=2.0}{m1,v3}
        \fmf{plain,rubout}{v2,v3}
      \end{fmfgraph*}}
    =
    \parbox{40pt}{
      \begin{fmfgraph*}(40,30)
        \fmfstyle
        \fmfcurved
        \fmfforce{(0.5w,-0.05h)}{c1}
        \fmfforce{(0.5w,1.05h)}{c2}
        \fmf{zigzag,w=2.0,f=(0.7,,0.7,,0.7)}{c1,c2}
        \fmfleft{i2,i1}
        \fmfright{o2,o1}
        \fmf{plain_arrow,t=4.0}{i1,v1}
        \fmf{plain,t=4.0}{i2,v1}
        \fmf{plain_arrow,t=4.0}{v2,o1}
        \fmf{plain,t=4.0}{v3,o2}
        \fmf{dbl_plain,t=0.0,rubout}{v1,m1}
        \fmf{plain,rubout}{v1,v2}
        \fmf{plain,rubout}{v1,v3}
        \fmf{plain}{v2,m1}
        \fmf{plain}{m1,v3}
      \end{fmfgraph*}}
    + \; c_1
    \parbox{40pt}{
      \begin{fmfgraph*}(40,30)
        \fmfstyle
        \fmfcurved
        \fmfforce{(0.5w,-0.05h)}{c1}
        \fmfforce{(0.5w,1.05h)}{c2}
        \fmf{zigzag,w=2.0,f=(0.7,,0.7,,0.7)}{c1,c2}
        \fmfleft{i2,i1}
        \fmfright{o2,o1}
        \fmf{plain_arrow,t=4.0}{i1,v1}
        \fmf{plain,t=4.0}{i2,v3}
        \fmf{plain_arrow,t=4.0}{v2,o1}
        \fmf{plain,t=4.0}{v2,o2}
        \fmf{plain,rubout}{v1,v2}
        \fmf{plain,rubout}{v3,v2}
        \fmffreeze
        \fmf{dbl_plain,r=0.5,rubout}{v3,v2}
        \fmf{plain}{v1,v3}
        \fmffreeze
        \fmf{phantom}{v1,m1,v3}
        \fmfv{d.sh=cross,d.si=10.0}{m1}
      \end{fmfgraph*}}
    + \; c_2
    \parbox{40pt}{
      \begin{fmfgraph*}(40,30)
        \fmfstyle
        \fmfcurved
        \fmfforce{(0.5w,-0.05h)}{c1}
        \fmfforce{(0.5w,1.05h)}{c2}
        \fmf{zigzag,w=2.0,f=(0.7,,0.7,,0.7)}{c1,c2}
        \fmfleft{i2,i1}
        \fmfright{o2,o1}
        \fmf{plain_arrow,t=4.0}{i1,v1}
        \fmf{plain,t=4.0}{i2,v1}
        \fmf{plain_arrow,t=4.0}{v2,o1}
        \fmf{plain,t=4.0}{v2,o2}
        \fmf{plain,l=1.0,rubout}{v1,v2}
        \fmf{plain,rubout}{v1,v2}
        \fmf{dbl_plain,r=1.0,rubout}{v1,v2}
      \end{fmfgraph*}}.
  \end{fmffile}
\end{align}
Here, the integral from the left-hand side stems from the fitfth and the
first integral on the right-hand side from the ninth topology of
Fig.~\ref{fig:tops}, and the coefficients $c_i$ are rational functions
in the kinematic invariants $m_H^2$ and $s$ and the space-time dimension
$D$.

\section{Conclusion}

The package {\tt TopoID} is a generic, process independent tool for
multiloop calulations, useful especially when many topologies are
involved.  Until now it has been applied successfully to the
$qq^\prime$-channel in Higgs boson production at
N$^3$LO~\cite{Anzai:2015wma} and to the NNLO soft-virtual corrections to
Higgs boson pair production~\cite{Grigo:2015dia}.  It can also handle
the classification of the massless five-loop propagator topologies,
including all their symmetries.  Along with this proceedings
contribution, a first public version can be obtained from the web page
\url{https://www.ttp.kit.edu/~jens/} where also future versions, a
detailed documentation and additional material will be provided.

\section*{Acknowledgments}

We would like to thank Alexey Pak who is the author of the initial
version of {\tt TopoID} and Matthias Steinhauser for his help related to
debugging the package.  This work was supported partially by the
European Commission through contract PITN-GA-2012-316704 (HIGGSTOOLS).
The attendence at ACAT was enabled by the generous support from the
DAAD~(German Academic Exchange Service).

\section*{References}

\bibliography{topoid-proc}
\bibliographystyle{iopart-num}

\end{document}